\documentclass[twocolumn]{aastex63}
\pdfoutput=1
\usepackage{xcolor}
\usepackage{makecell}
\usepackage{pbox}

\shorttitle{Turbulence in Galactic Centers}
\shortauthors{Salas et al.}

\newcommand{\x}{\color{blue}}

\newcommand{\msun}{M$_{\odot}$}
\newcommand{\gadget}{Gadget2}
\newcommand{\e}[1]{$\times10^{#1}$}

\newcommand{\invcmcube}{cm$^{-3}$}

\newcommand{\nt}{$N_t$}
\newcommand{\eturb}{$\Delta E_{in}$}
\newcommand{\splash}{\texttt{SPLASH}}

\begin{document}

\title{Modeling Turbulence in Galactic Centers}

\correspondingauthor{Jesus~M.~Salas}
\email{jesusms@astro.ucla.edu}

\author{Jesus~M.~Salas}
\affiliation{Dept. of Physics \& Astronomy, University of California, Los Angeles, CA, 90095, USA}

\author{Mark~R.~Morris}
\affiliation{Dept. of Physics \& Astronomy, University of California, Los Angeles, CA, 90095, USA}

\author{Smadar~Naoz}
\affiliation{Dept. of Physics \& Astronomy, University of California, Los Angeles, CA, 90095, USA}
\affiliation{Mani L. Bhaumik Institute for Theoretical Physics, Department of Physics and Astronomy, University of California, Los Angeles, CA 90095, USA}

\begin{abstract}
Turbulence is a prevalent phenomenon in the interstellar medium, and in particular, the environment at the centers of galaxies. For example, detailed observations of the Milky Way's Central Molecular Zone (CMZ) revealed that it has a complex and turbulent structure. Turbulence on galactic scales is often modeled using star formation and feedback. However, these effects do not appear to be sufficient for explaining the high-velocity dispersion observed in the CMZ, indicating that additional gas-stirring processes are likely to be operating. Here we introduce a proof-of-concept method to drive turbulence in gas that orbits under the influence of a galactic potential. Instead of relying on a particular physical mechanism, we have adopted a Fourier forcing module and have applied it using a Smoothed Particle Hydrodynamics code. To test our method, we performed simulations of a simplistic model of the CMZ. 
Our turbulence injection method is capable of balancing the self-gravity of the gas, which allows us to run the simulations for long timescales and thereby follow the evolution of the CMZ. Our results show that turbulence induces a flocculent spiral pattern in our model, analogous to that found in galactic-scale simulations. Furthermore, we find that our turbulence injection method induces inward migration of gas, a result consistent with previous numerical simulations. We submit that this injection method is a promising new tool to simulate turbulence in galactic centers. 
\end{abstract}


\section{Introduction}

Turbulence is one of the major processes that governs
the structure and evolution of the interstellar medium \citep[ISM,][]{Elmegreen2004ARAA..42..211E}. For example, supersonic  turbulence  is  known  to be  a dominant  process in regulating (both inhibiting  and  fostering) star formation in molecular clouds \citep{McKee2007ARAA..45..565M}. However, galactic
centers show a number of interesting deviations from the star formation behavior seen at larger galactic radii. Molecular clouds in our own Galactic Center (GC) show systematic differences in their properties from
disk clouds: they appear to have thermal, turbulent and magnetic pressures much higher than those present in the large-scale Galactic disk \citep[e.g.,][]{Spergel1992}. 

The main gaseous feature of the GC, the Central Molecular Zone (CMZ), has a rich and complex structure that extends over a galactocentric radius of $\sim300$ pc and contains a mass of M$\sim3-7$\e{7} \msun\ \citep[e.g.,][]{Morris1996}. 
It is largely composed of relatively dense ($n\sim10^3$-$10^5$ \invcmcube) and warm gas ($\sim70-100$ K on average, e.g., \citealt{Gusten1981A&A...103..197G,Morris1983AJ.....88.1228M,Huettemeister1993A&A...280..255H,Ao2013A&A...550A.135A}), mostly condensed into Giant Molecular Clouds (GMCs) or dense tidal streams of molecular gas. These relatively warm gas temperatures are one of the key properties of CMZ clouds, and there is evidence showing that the gas is kept warm by the dissipation of turbulence \citep[e.g.,][]{Immer2016, Ginsburg2016}. It has also been suggested that turbulence plays a role in the suppression of star formation in this region (e.g., \citealt{Kruijssen2014MNRAS.440.3370K}). However, the driving mechanism for the turbulence in CMZ clouds has not been conclusively identified (see \citealt{Kruijssen2014MNRAS.440.3370K} for a discussion of possible sources of turbulence). Furthermore, the large turbulent velocity dispersion within the CMZ must be responsible for supporting the gas against gravitational collapse, since the thermal pressure of the gas would be insufficient. 

Interstellar turbulence decays quite rapidly, on timescales of the order of the free-fall time of the system \citep[e.g.,][]{MacLow1999}. Therefore, energy must be injected into the system in order to maintain the turbulence. Simulations of turbulence-driven gas are often 
employed in studies of the interstellar medium and star formation \citep[e.g.,][]{Stone1998ApJ...508L..99S,MacLow1998PhRvL..80.2754M,Krumholz2005ApJ...630..250K,Burkhart2009ApJ...693..250B,Federrath2010AA...512A..81F}. Typically, this is achieved by a Fourier forcing module, which can be modeled with a spatially static pattern in which the amplitude is adjusted in time \citep{Stone1998ApJ...508L..99S,MacLow1999}. Other studies employ a forcing module that can vary both in time and space \citep[e.g.,][]{Padoan2004PhRvL..92s1102P,Schmidt2006,Federrath2010AA...512A..81F}.

In the case of galactic-scale simulations, driven turbulence is mimicked by injecting energy due to supernovae (SN). For example, simulations by \citet{Kim2011ApJ...735L..11K,Emsellem2015MNRAS.446.2468E,Shin2017ApJ...841...74S,Seo2019ApJ...872....5S,Armillotta2019MNRAS.490.4401A}, and \citet{Tress2020MNRAS.499.4455T} have modeled turbulence by using star formation and SN feedback models. In general, these models depend on underlying assumptions regarding star formation rates, SN energies and injection rates. Furthermore, recent studies \citep[e.g.,][]{Scannapieco2012MNRAS.423.1726S,Rosdahl2017MNRAS.466...11R,Keller2020arXiv200403608K} have demonstrated that the different choices of SN feedback model (including the underlying physical processes driving the feedback) produce significant differences in morphology, density, etc, of the simulated galaxies. 

Here we introduce a proof-of-concept method to drive turbulence in gas that orbits under the influence of a galactic potential. Instead of relying on a particular physical mechanism, we adopt a Fourier forcing module, which has the advantage of being independent of the source of turbulence. Our turbulence treatment is based on the method described by \cite{MacLow1999}, in which a turbulent velocity field is drawn from a spatially static pattern having a power spectrum $P(k)\propto k^{-n}$, where $k$ is the wavenumber. We apply our method to a smoothed particle hydrodynamics (SPH) code, and we test its effectiveness using a simplistic model of the CMZ. Our simulations consider self-gravity (i.e., the mutual gravitational interactions between the SPH particles) and the effects of pressure from a surrounding medium.

This paper is organized as follows: Section \ref{sec:numerical_methods} summarizes the numerical methods, with further details on our turbulence method in Appendix \ref{apen:derivation}. Section \ref{sec:conv} and Appendix \ref{apen:convergence} describe the tests performed to verify the effectiveness of our turbulence method. We present our main results in Section \ref{sec:results}, and conclude in Section \ref{sec:discussion}.

\section{Numerical Methods}\label{sec:numerical_methods}

We used the N-body/SPH code \gadget\ \citep{Springel2005}, which is based on the tree-Particle Mesh method for computing gravitational forces and on the SPH method for solving the Euler equations of hydrodynamics. The smoothing length of each particle in the gas is fully adaptive down to a set minimum of 0.001 pc. \gadget\ employs an entropy formulation of SPH, as outlined in \cite{Springel2002}, with the smoothing lengths defined to ensure a fixed mass (i.e., fixed number of particles) within the smoothing kernel volume (set for N$_{neigh}$ = 64). The code adopts the Monaghan-Balsara form of artificial viscosity \citep{Monaghan1983,Balsara1995}, which is regulated by the parameter $\alpha_{MB}$, set to 0.75.

We modified the standard version of \gadget\ to include turbulence driving, the gravitational potential of a Milky Way-like galaxy, and the effects of pressure by a surrounding medium. We describe these modifications below. 

\subsection{External Pressure}\label{subsec:pressure}
\begin{figure*}
	\centering
	\includegraphics[width=\textwidth]{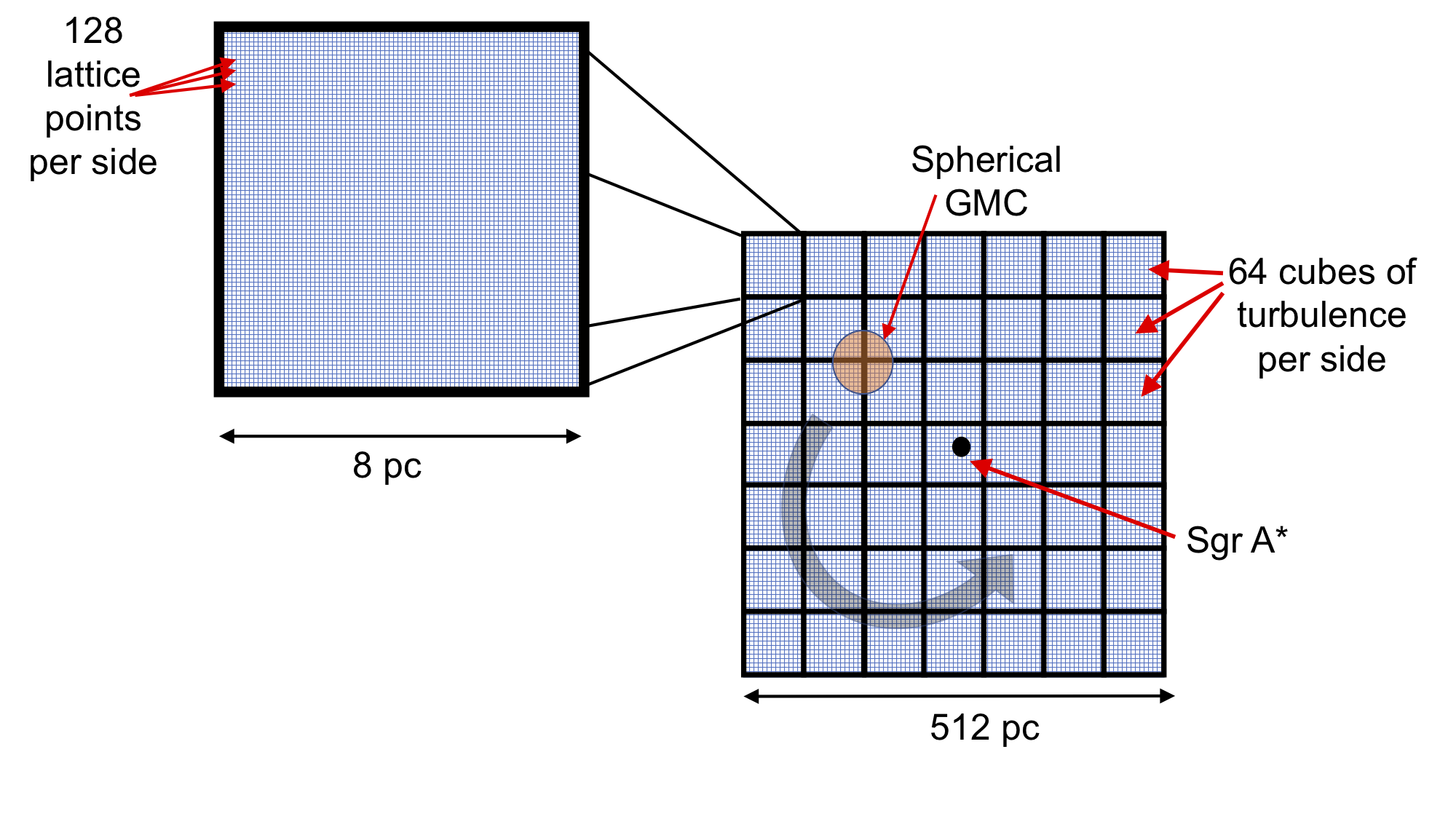}
	\caption{Two dimensional representation of our turbulence driving method. A single turbulence cube is shown on the left, while the combination of cubes to fill up the large simulation box is shown on the right. Note that the above schematic is intended for illustration purposes, since the number of cubes sketched is different from what we actually used.}
		\label{fig:schematic}
\end{figure*}
The interstellar medium of the GC is modeled via an external pressure term to approximate a constant pressure boundary.  
Following \cite{Clark2011}, we modify \gadget's momentum equation \citep{Springel2002}:
\begin{equation}\label{eq:sph_motion2}
\frac{d v_i}{dt} = - \sum_{j} m_j \left[  f_i \frac{P_i}{\rho^2_i} \nabla_i W_{ij}(h_i) + f_j \frac{P_j}{\rho^2_j} \nabla_i W_{ij}(h_j)  \right]
\end{equation} 
where $v_i$ is the velocity of particle $i$, $m_j$ is the mass of particle $j$, $P_i$ is the pressure, $\rho_i$ is the density, $W_{ij}(h_i)$ is the kernel function which depends on the smoothing length $h_i$, and $f$ is a unitless coefficient that depends on $\rho_i$ and $h_i$. We replace $P_i$ and $P_j$ with $P_i - P_{ext}$ and $P_j - P_{ext}$, respectively, where $P_{ext}$ is the external pressure. The pair-wise nature of the force summation over the SPH neighbors ensures that $P_{ext}$ cancels for particles that are surrounded by other particles. At the boundary, where the $P_{ext}$ term does not disappear, it mimics the pressure contribution from a surrounding medium \citep{Clark2011}. We set $P_{ext}$ equal to $10^{-10}$ ergs \invcmcube, an approximate value for the GC \citep{Spergel1992,Morris1996}. 
\subsection{The galactic potential} \label{subsec:potential}
The gravitational potential we use is adopted from \cite{Zhao1994AJ....108.2154Z}, which is a modified version of the prolate bar potential introduced by \cite{Binney1991MNRAS.252..210B}. This potential has the form: 
\begin{equation}\label{eq:potential}
\Phi(r,\theta,\phi) = 4\pi G \rho_0 r_0^2 \left(\frac{r}{r_0}\right)^\alpha P(\theta,\phi) \ ,
\end{equation}
where $(r,\theta,\phi)$ are spherical coordinates fixed on the rotating bar ($r = \sqrt{x^2+y^2+z^2}$, where $x,y,z$ are the standard Cartesian coordinates. The supermassive black hole would be at $r = 0$, the bar's major axis is aligned with the $x$-axis, and the $z$ axis represents the vertical direction, with the galactic plane at $z=0$), and $P(\theta,\phi)$ is the associated Legendre function, which can be written as:
\begin{equation}
P(\theta,\phi) = \frac{1}{\alpha (1+\alpha)} - \frac{Y(\theta,\phi)}{(2-\alpha)(3+\alpha)} \ .
\end{equation}
$Y$ is a linear combination of spherical harmonic functions of the
$l= 2$, $m= 0,2$ modes:
\begin{equation}
Y(\theta,\phi) = -b_{20} P_{20} (\cos\theta) + b_{22}P_{22}(\cos\theta)\cos2\phi \ .
\end{equation}
The parameter $b_{20}$ determines the degree of oblateness/prolateness
while $b_{22}$ determines the degree of non-axisymmetry. Motivated by the previous work of \cite{Kim2011ApJ...735L..11K}, and more recently of \cite{Gallego2017}\footnote{We note that there is a negative sign misprint in \cite{Kim2011ApJ...735L..11K} (their Equation 2) and in \cite{Gallego2017} (their associated Legendre function).}, we use the parameters: $\alpha = 0.25$, $b_{20}=0.3$, $b_{22} =0.1$, $\rho_0 =40$ \msun\ pc$^{-3}$ and $r_0 = 100$ pc. Given these parameters, a bar with axis ratios of [1: 0.74: 0.65] is obtained for the isodensity surface that intersects points [$x$ = 0, $y$ = $\pm200$~pc, $z$ = 0].
Enclosed masses inside $200$~pc and $1000$~pc are $10^9$ \msun\ and $7\times10^9$ \msun, respectively.

In addition to the gravitational force due to the potential
above, we introduced the rotation of the bar by adding centrifugal and Coriolis forces. We used the most recent estimate for the Galactic bar's pattern speed ($\Omega_{bar} = 40$ km s$^{-1}$ kpc$^{-1}$; e.g., \citealt{Hawthorn2016ARAA..54..529B,Portail2017MNRAS.465.1621P}).

\subsection{Turbulence Driving}\label{subsec:turbulence}
As noted above, driven turbulence is often modeled using one of two types of Fourier forcing modules. Both methods require Fourier transforms on a cubic grid (or lattice) with $N^3$ points (or a square lattice with $N^2$ on 2D simulations, where typical values for $N$ = 128, 256, 1024, etc). By imposing this cubic lattice onto a simulation box with a physical size of $L$ per side, we can use the separation between adjacent lattice points ($L/N$) as a proxy for the resolution of the turbulence. Hence, for a large-scale simulation such as the CMZ environment, e.g., $L = 500$~pc, and simultaneously resolving turbulence on small scales, e.g., $0.01$ pc, the turbulence cubic lattice would have to contain $N^3$ = $5000^3$ points, which would require massive computational resources. To circumvent this limitation, we instead use many smaller turbulence grids to fill the volume of our larger simulation box, as follows:

First, we create a library of 10 files, which our modified version of the \gadget\ code reads in at the start of the simulation. Each file contains a unique realization of a turbulent velocity field (in the form of a 3D matrix) with power spectrum $P(k) \propto k^{-4}$ (suitable for compressible gas; e.g., \citealt{Clark2011}). Each of these 3D matrices of turbulence is generated using the methods described in \cite{Rogallo1981} and \cite{Dubinski1995}: via fast Fourier transforms inside a $128^3$ box. 

\begin{figure}
	\centering
\includegraphics[width=0.5\textwidth]{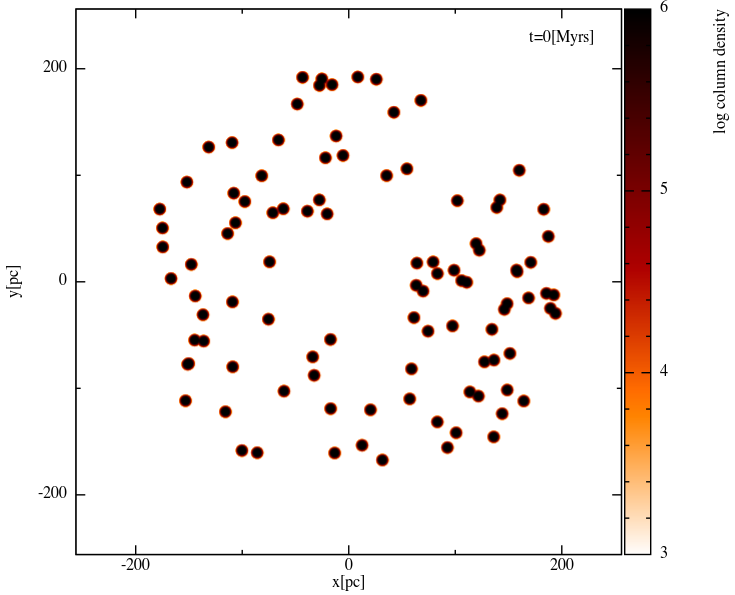}
\includegraphics[width=0.5\textwidth]{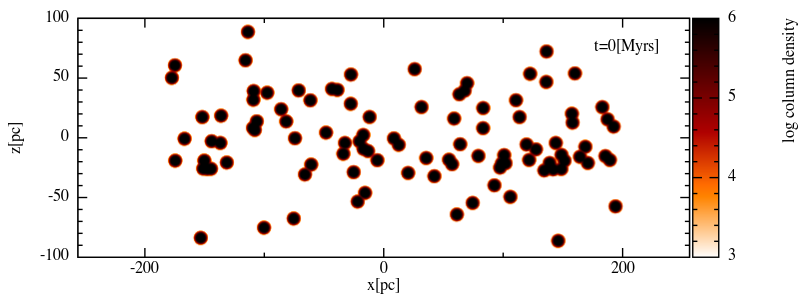}
\caption{Initial conditions for our simulations, x-y plane and edge-on view. Color bar indicates column number density in log scale (units in \invcmcube\ pc).}
\label{fig:ICs}
\end{figure}

These 3D turbulence matrices can be visualized as lattice cubes (or grids) with equally spaced lattice points, containing $128\times128\times128$ points. The parameter that sets the physical size of these cubes is named $L_{cube}$. In the test model used in this paper, we make the simplifying assumption that turbulence is injected on scales similar to the average size of CMZ clouds, i.e., $L_{cube}$ = 8 pc per side. Thus, the separation between two adjacent lattice points along one axis is $8$pc$/128 = 0.0625$ pc. 

In some studies, the driving module only contains power on the larger scales (e.g., \citealt{Federrath2010AA...512A..81F}). This type of driving models the kinetic energy input from large-scale turbulent fluctuations, which then break up into smaller structures as the kinetic energy cascades down to scales smaller than the turbulence injection scale. However, in SPH, the artificial viscosity may damp this energy cascade and prevent it from reaching the smaller scales. Consequently, to create the different realizations of turbulent velocity fields, we use a discrete range of $k$ values from $k_{min} = 2$ to $k_{max} = 128$, thus effectively injecting energy on scales between $L_{cube}/2 = 4$~pc (for $k=2$) and $L_{cube}/128 = 0.06$~pc (for $k=128$). To create the initial (turbulent) velocity field of an individual cloud (see Section \ref{subsec:IC}), we use trilinear interpolation to calculate the velocity components for each SPH particle, based on that particle's position on a turbulence cube. This interpolation method results in a turbulent velocity function $\vec{I}(x,y,z)$.  

Next, we use $64\times64\times64$ cubes of turbulence to fill up the volume of our large simulation box. This gives us a simulation box of size $L_{global}$ = $8$pc$\times64 = 512$ pc per side. Thus, the spatial resolution of the turbulence in our large simulation box is the same as the resolution of an individual turbulence lattice cube. Each of the turbulence lattice cubes that fill up the large simulation box is randomly chosen from our library of 10 files, thereby avoiding a velocity field that is coherent over scales greater than $L_{cube}$. A 2D graphic representation of the method we describe here is shown in Figure \ref{fig:schematic}. This method raises a concern regarding the interface between the turbulence cubes, namely, that there will be discontinuities in the turbulent velocity kicks throughout the gas at the cubes' interfaces. However, the turbulent velocity added to the gas is quite small in comparison with the orbital velocity, and indeed, we find in practice that there are no obvious shocks or discontinuities at the interfaces that exceed those induced by the injected velocity increments (see Section \ref{sec:results}). 

To drive the turbulence, we follow a method similar to that described by \cite{MacLow1999}: every $N_t$ timesteps (we fixed the timestep in all our simulations to $\Delta t_s = 1000$ yrs) we add a velocity increment to every SPH particle, given by:
\begin{equation}
\label{eq:kicks}
\Delta \vec{v}(x,y,z) = A \sqrt{G \rho} \text{     }  \vec{I}(x,y,z)\ ,
\end{equation}
where $\vec{I}(x,y,z)$ is the turbulent velocity interpolated from the turbulence field of the cube that contains the particle in question. The amplitude $A$ is chosen to maintain a constant kinetic energy input rate $\dot{E_{in}} = \Delta E_{in}/(N_t \Delta t_s)$, and the term $\sqrt{G \rho}$ is added to counteract gravitational collapse by ensuring that higher-density regions receive more kinetic energy\footnote{This density factor stems from the assumption that molecular clouds are supported against gravitational collapse by turbulence, and therefore the velocity kicks should be inversely proportional to the free-fall time,  $t_{ff} \propto (G\rho)^{-1/2}$.}. Any particle outside the ($512$~pc)$^3$ simulation box does not receive any turbulent energy. 

For compressible gas with a time-dependent density distribution, maintaining a constant kinetic energy input rate requires solving a quadratic equation in the amplitude $A$ every time the driving is performed \citep{MacLow1999}. For $N_{sph}$ particles of mass $m_p$, each with density $\rho_i$, $A$ is derived from (see Appendix \ref{apen:derivation}):
\begin{equation}
\label{eq:A}
\Delta E_{in} = \frac{1}{2} m_p \sum_{i=1}^{N_{sph}} \left[      A^2  G \rho_i \vec{I_i} \cdot \vec{I_i}  + 2 A \sqrt{G \rho_i} \text{     }  \vec{I_i} \cdot \vec{v_{1,i}}        \right]\ .
\end{equation}
Following \cite{MacLow1999}, we take the larger root as the value for $A$. 

Therefore, our methodology introduces two adjustable parameters: \eturb, the energy input per injection, and \nt, the number of timesteps between injections.

Finally, to mimic the random nature of turbulence, we change the turbulent velocity field of each of the $64^3$ cubes every time the driving is performed. This is done by replacing  each of the cubes with a different one, chosen randomly from the 10 files in our library. See Appendix \ref{apen:derivation} for further details.

\subsection{Initial conditions}\label{subsec:IC}
To initiate the computations, we use a fiducial model of the central region of the Milky Way, consisting of a collection of $100$ initially isolated, spherical GMCs. Each cloud contains $N_p=10^4$ particles, with mass $m_p$ = $30$ \msun\ per particle. The clouds are distributed randomly in an annular disk of inner radius $30$ pc, outer radius $200$ pc, and a Gaussian scale height of $30$ pc. Each cloud has a radius of $4$ pc, and an initial turbulent velocity field such that $|E_{turb}/E_{pot}| = 0.5$ (i.e., they are initially in virial equilibrium). The clouds' initial center of mass velocities were set so that they move on circular orbits and are parallel to the galactic plane, with their magnitudes ($v_{\phi})$ calculated using the potential described in Section \ref{subsec:potential}. In order to also give the system a broader vertical structure, the initial $v_z$ components of each cloud were set such that $v_z$ = 0.5$v_\phi$, with the $v_z$ vector always pointing towards the galactic plane. Figure \ref{fig:ICs} shows the initial snapshot of our CMZ model.

All simulations were run using an isothermal equation of state with $T$ = $100$~K. This assumption of isothermal gas is somewhat crude, but may still provide an adequate physical approximation to the real thermodynamics in dense molecular gas \citep{Wolfire1995,Pavlovski2006}.  

During our testing phase, we also considered an initially uniform disk as our initial condition. Except for the time to reach a steady-state (see Section \ref{sec:results}), the end result was qualitatively identical to the model described in this paper. We therefore conclude that the specific features of the initial conditions are unimportant for our purposes as long as the particles are initially distributed over the same domain. We opted for a collection of clouds as our initial conditions due to the flexibility in setting the position, size, mass, initial turbulent velocity field, etc, of each individual cloud.

\section{Convergence and consistency tests}\label{sec:conv}   
When the self-gravity of the gas is included, combined with a relatively low temperature, our CMZ simulations without injected turbulence experience runaway gravitational collapse, causing the simulations to fail within a dynamical timescale ($\sim0.1-0.3$~Myrs).  This effect can be alleviated by increasing the softening length (which in our simulations is set equal to the smoothing length, i.e., 0.001 pc). In such a case, the simulation can run for long timescales, however the gas concentrates to unphysically large densities, causing the simulation to slow down to an impractical pace, and is thus an expensive use of computing resources. This result thereby emphasizes the importance of the turbulence injection method introduced in this paper. However, turbulence injected too infrequently leads to the same result as if there was no turbulence injection: gas collapses locally to unphysically large densities (see Appendix \ref{apen:convergence}). Therefore, it is important to inject turbulence relatively often. 

In practice, the turbulence driving method described in Section \ref{subsec:turbulence}, adds an additional velocity ``kick'' to each particle every \nt\ timesteps. The energy \eturb\ injected per \nt\ timestep is kept constant. With these two free parameters we conducted several tests using our CMZ model to find the optimal range of values that give rise to reasonable densities over long timescales. We varied the energy input \eturb\ from $10^{46}$ to $10^{50}$ ergs, and $N_t$ from $1$ to $5$. We describe in detail all performed tests in Appendix \ref{apen:convergence}. Our tests led us to the choice of parameters \nt\ = $2$, \eturb\ = $10^{47}$~ergs (e.g., see Figure \ref{fig:convergence2}).  

\section{Results} \label{sec:results}
\begin{figure*}
\centering
\includegraphics[width=1\textwidth]{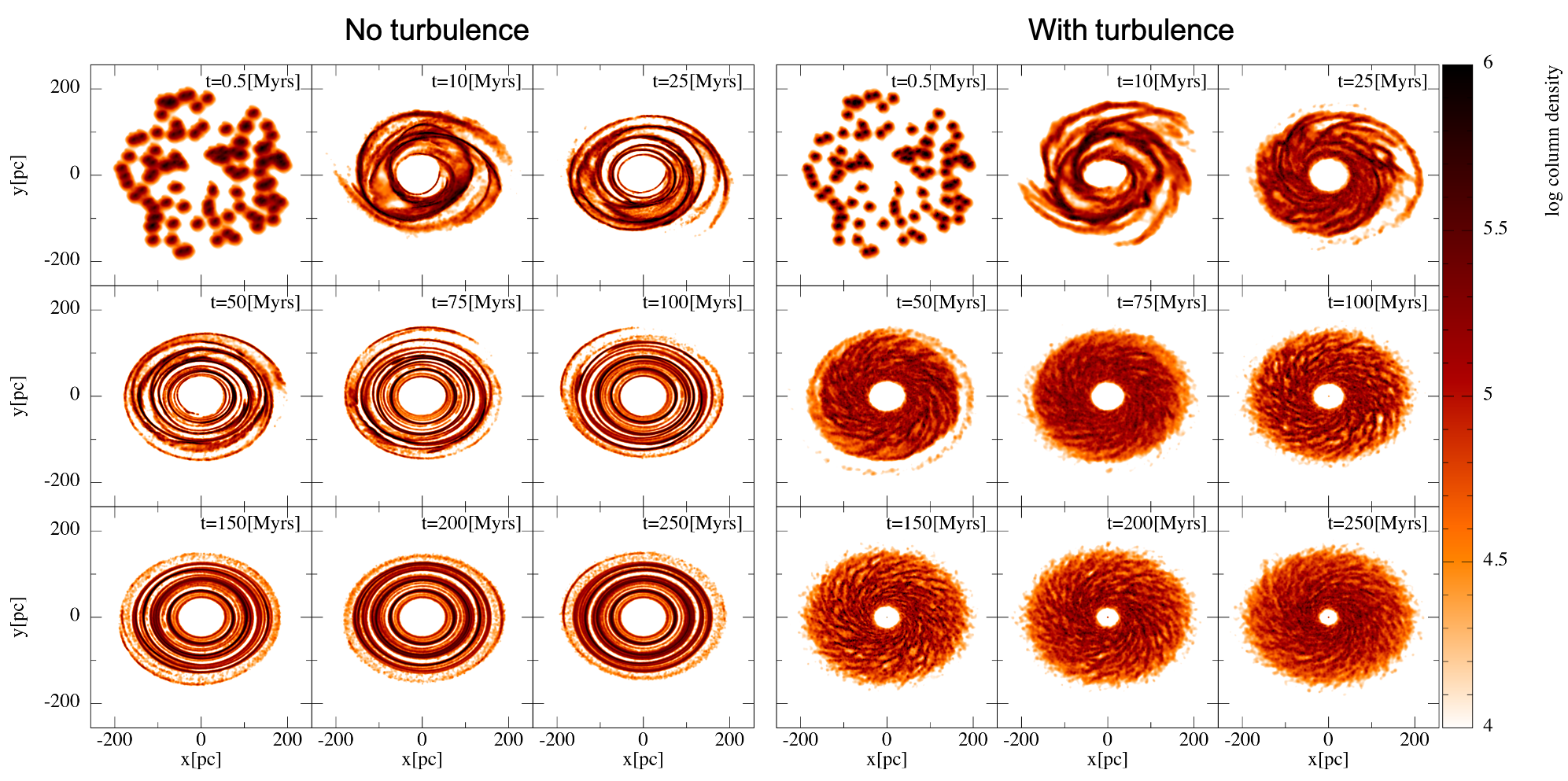}
\caption{Column number density at different times for the no-turbulence run (left) and turbulence run (right).
The long axis of the bar is oriented along the x-axis. The color map is in logarithmic scale. Units are in \invcmcube\ pc.}
\label{fig:xy}
\end{figure*}

\begin{figure*}
	\centering
	\includegraphics[width=\textwidth]{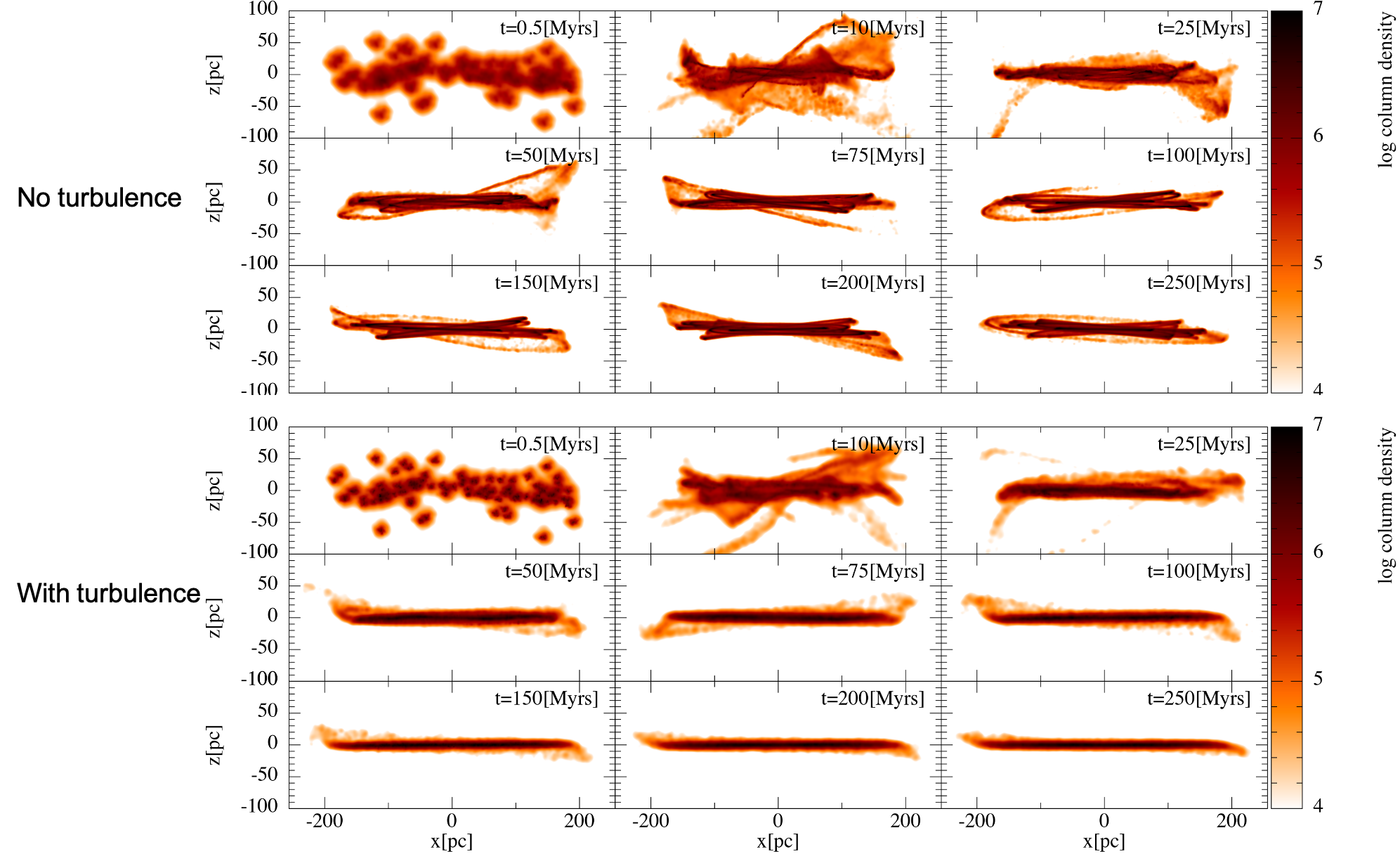}
	\caption{Edge-on view of the column number density at different times for the no-turbulence run (top) and turbulence run (bottom).  The long axis of the bar is oriented along the x-axis. The color map is in log scale. Units are in \invcmcube\ pc. The simulation without turbulence exhibits a more pronounced vertical structure than the simulation with turbulence. This is because in the simulation with turbulence, self-gravity is included, and thus the mutual gravitational interactions between particles compresses the disk to the galactic plane. } 
	\label{fig:zx}
\end{figure*}
We ran the system for $250$~Myrs, using the initial conditions described in Section \ref{subsec:IC}, the turbulence parameters described in Section \ref{sec:conv}, and including self-gravity. The system reaches a semi-steady state after $50-100$~Myrs regardless of the turbulence parameters (as described in Appendix \ref{apen:convergence}),  and thus the choice of $250$~Myrs allows us to capture the relevant dynamics (see below). We also ran a simulation without turbulence (and self-gravity turned off) for comparison. We show the face-on and edge-on views of the resulting column density evolution of our simulations in Figures \ref{fig:xy} and \ref{fig:zx}, respectively. The results shown here indicate that our turbulence injection module is capable of balancing the self-gravity of gas concentrations, which allows us to run the simulation for long timescales. Furthermore, discontinuities due to the grids' interfaces are unnoticeable, as anticipated. 

In both runs, the clouds are tidally stretched relatively quickly, and the gas settles into a disk after $\sim50$~Myrs. In the simulation without turbulence, the clouds are stretched into gas streams which comprise the disk, and reaches steady state by $\sim50$~Myrs. The inner inner cavity of radius $\sim30$~pc, which is a product of the initial conditions, remains unchanged for the entire simulation. 

However, there are two major differences between the two simulations. First, the resulting streams in the simulation with turbulence coalesce into a disk with a flocculent spiral pattern. Second, the inner cavity fills in slowly over time. We address each of these differences below.

\subsection{Spiral structure}
After running for $\sim50$~Myrs, the gas in the run with turbulence settles into a quasi-steady state with a flocculent spiral pattern (because of the constantly injected turbulence, a perfect steady state cannot occur). These spiral segments are attributable to the dynamical response of our self-gravitating, shearing disk to local density perturbations (e.g., \citealt{Julian1966ApJ...146..810J}), which in this case are caused by the forced turbulence. It is a common result that turbulence promotes the development of high-density regions due to convergent flows (e.g., \citealt{Elmegreen2004ARAA..42..211E,McKee2007ARAA..45..565M}). 

A similar result was found on galactic scales by \cite{Donghia2013ApJ...766...34D}. The spiral patterns in their simulations of self-gravitating disks of stars are not global as predicted by classical static density wave theory, but locally they appear to fluctuate in amplitude with time. Their spirals are actually segments produced by sheared local under-dense and over-dense regions. These under-dense and over-dense regions act as gravitational perturbers, maintaining the local spiral morphology. In our case, it is the injected turbulence that acts as the local perturber in the gas.

Furthermore, observations of galactic centers have revealed intricate dust structures that are often organized in a clear spiral pattern. For example, the survey studies by \cite{Regan1999AJ....117.2676R,Martini1999AJ....118.2646M,Pogge2002ApJ...569..624P} and \cite{Martini2003aApJS..146..353M,Martini2003bApJ...589..774M} indicate that $\sim50-80$\% of galaxies in their samples possess nuclear spirals, regardless of their nuclear activity.  The  pattern  of  some of the observed nuclear spirals is highly organized, similar to grand-design spirals in main galactic discs. Others display a more chaotic, or flocculent, spiral pattern. A direct comparison between our simulations and observations is difficult due to the limited spatial resolution of the observations, however, our results suggest that turbulence could be (at least partially) responsible for the spiral pattern observed in the gas layer in the centers of galaxies.  

\subsubsection{Effects of external pressure}
\begin{figure*}
    \centering
    \includegraphics[width =1\textwidth]{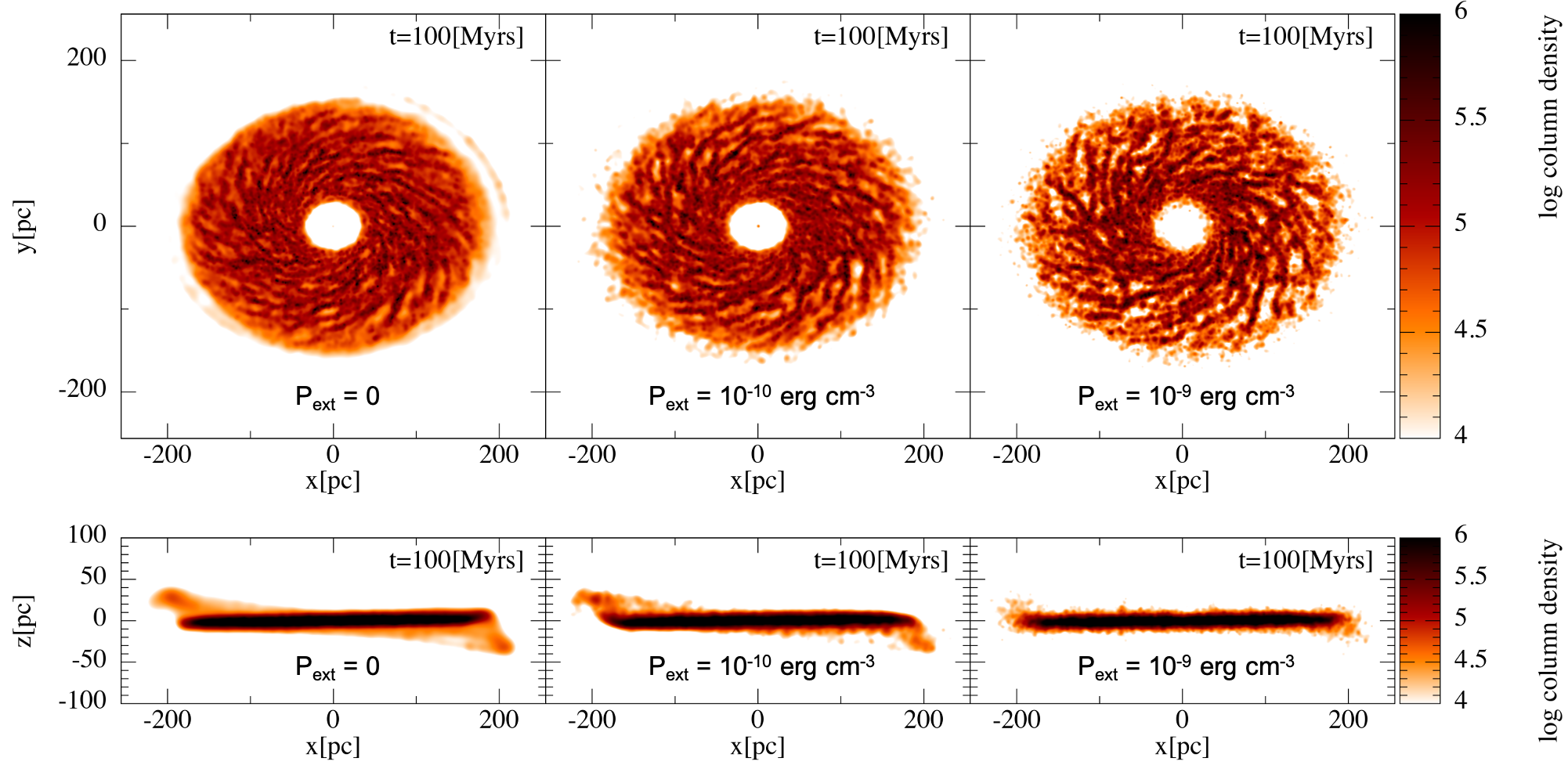}
    \caption{  Column number density maps of three simulations performed to compare between different values of $P_{ext}$. The turbulence parameters are the same as our fiducial run (\nt=2, \eturb=$10^{47}$ ergs). The top panels show the face-on views, while the bottom panels show the edge-on views. Left panels: vacuum boundaries (no external pressure). Middle panels: $P_{ext}$ = $10^{-10}$ erg cm$^{-3}$, same as our fiducial model (e.g., Figure \ref{fig:xy}). Right panels: $P_{ext}$ = $10^{-9}$ erg cm$^{-3}$, i.e., a factor of 10 greater than our fiducial model. 
    The external pressure pushes the lower density gas into the disk and enhances the density contrast of the gas spiral arms. }
    \label{fig:pressure_compare}
\end{figure*}

As discussed in Section \ref{subsec:pressure}, we modified Gadget2 to model an external pressure boundary, as opposed to the vacuum or periodic boundary conditions that are the only choices in the standard version of the code.  We used an observationally motivated value of $P_{ext}$ = $10^{-10}$ erg cm$^{-3}$ for the Galactic Center. However, it is worth investigating how varying this pressure term affects the gas morphology. We ran two additional simulations with $P_{ext}$ = 0 (standard vacuum boundary conditions) and $P_{ext}$ = $10^{-9}$ erg cm$^{-3}$, which is an order of magnitude greater than the value from our fiducial simulations. We show the column density maps (both face-on and edge-on) at $t$ = 100 Myr in Figure \ref{fig:pressure_compare}. As depicted, without the external pressure term (see left panels in Figure \ref{fig:pressure_compare}), the gas disk has a smooth boundary at the edges, as well as material above and below the plane. When the external pressure term is introduced with our fiducial value, it pushes the low density gas at the edge towards the disk, and it increases the density contrast of the flocculent spiral pattern. Similarly, when the pressure term is large ($P_{ext}$ = $10^{-9}$ erg cm$^{-3}$, right panels in Figure \ref{fig:pressure_compare}), the density contrast of the spirals is increased further. Furthermore, the material above and below the plane is pushed towards the disk. Interestingly, this behaviour suggests an observational test for the value of external pressure, which may be used for systems in which the pressure is undetermined. 

\subsection{Inward migration}

\begin{figure}
\centering
\includegraphics[width=.5\textwidth]{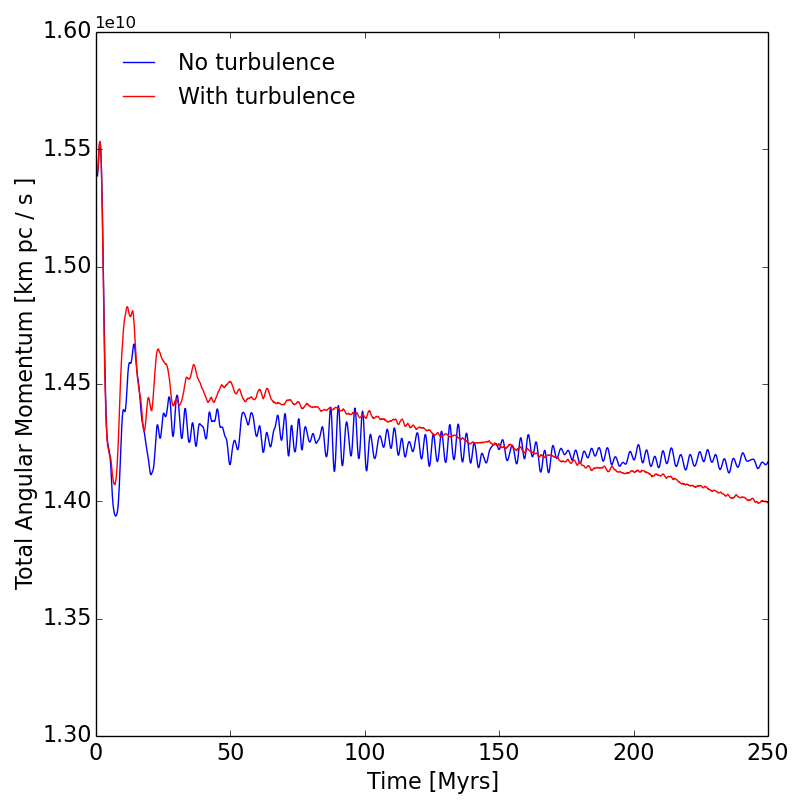}
\caption{Evolution of the total specific angular momentum ($l$) in the simulations. In the run with turbulence, $l$ decreases slightly, with a difference of $\sim3\%$ between 50 and 250 Myrs. However, this change is sufficiently small that the angular momentum can be considered approximately constant. }
\label{fig:angular}
\end{figure}

The inner cavity closes slowly over time, as gas migrates inward. This might imply a loss of angular momentum in the system due to the turbulence. In Figure \ref{fig:angular} we show the total (specific) angular momentum over time for both simulations. Despite our strategy for avoiding adding a net angular momentum to the simulations (see Appendix \ref{apen:derivation}), we see that, in the run with turbulence, angular momentum slightly decreases, from  $\sim1.45\times10^{10}$ km s$^{-1}$ pc at $50$~Myrs to $\sim1.4\times10^{10}$ km s$^{-1}$ pc at $250$~Myrs, a difference of $\sim3\%$. Clearly, the sign switching strategy we describe in Appendix \ref{apen:derivation} does not completely ensure angular momentum conservation. Furthermore, it has been shown that turbulence can impose a nonzero angular momentum \citep[e.g.,][]{Clark2011}. However, this change is quite small over long timescales, and thus we do not consider this to be the reason for the inward migration\footnote{ We verified that the inward migration is indeed a feature of the injected turbulence by considering the possibility that the interaction between turbulence and the rotating, non-axisymmetric potential could have led to this angular momentum decrease. We tested this idea by running a simulation with a rotating, axisymmetric potential (by setting $b_{22}$ = 0 and $\Omega_{bar}$ = 40 km s$^{-1}$ kpc$^{-1}$, see Section \ref{subsec:potential}), and another with a non-rotating, axisymmetric potential (by setting $b_{22}$ = 0 and $\Omega_{bar}$ = 0). However, in both cases, we recover the same negative slope as in our fiducial model (red line in Figure \ref{fig:angular}). Thus, we conclude that this small angular momentum reduction is a feature of the injected turbulence. }.

\begin{figure*}
\includegraphics[width=\textwidth]{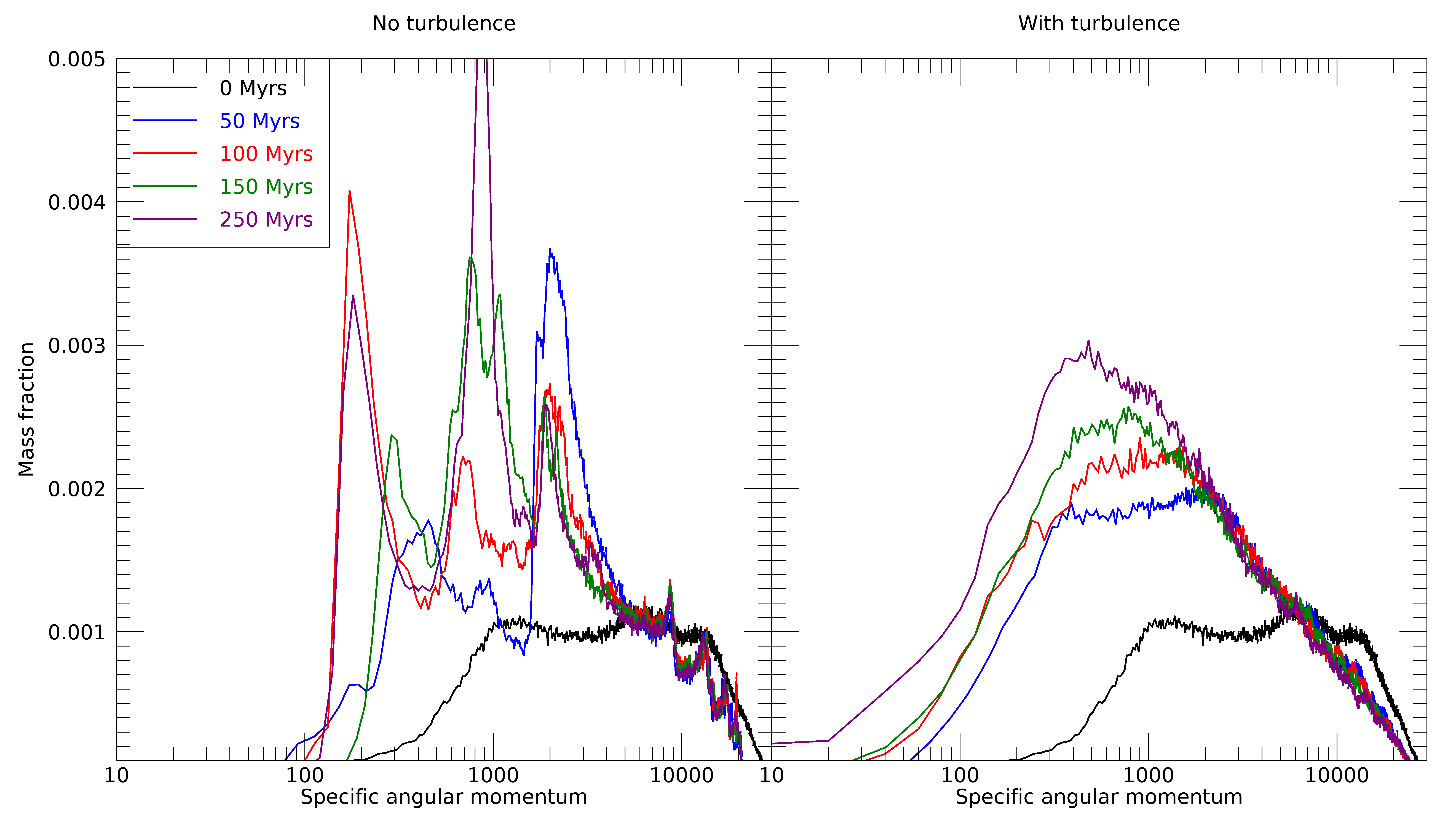}
\caption{Distribution of the mass fraction of gas as a function of specific angular momentum (units of km pc s$^{-1}$) at different times for the simulations without turbulence (left) and with turbulence (right). The graph was calculated using a specific angular momentum bin size of 20 km pc s$^{-1}$. In the simulation without turbulence, the shape of the distribution is relatively unchanged after 100 Myrs. In contrast, the angular momentum settles into a single-peak distribution in the run with turbulence, and this peaks shifts towards lower angular momentum values over time, thus accounting for the inward migration of gas. }
	\label{fig:angular_pdf}
\end{figure*}

Conversely, it has been shown that supersonic turbulence inside accretion disks (e.g., \citealt{Wang2009ApJ...701L...7W}) can promote accretion onto SMBHs by enhancing angular momentum transfer (e.g., \citealt{Collin2008A&A...477..419C,Chen2009ApJ...695L.130C}). In particular, \citet{Hobbs2011MNRAS.413.2633H} demonstrated using numerical simulations of supermassive black hole accretion that turbulence can broaden the angular momentum distribution, setting some gas on low angular momentum orbits. We find a similar result in Figure \ref{fig:angular_pdf}, where we show the mass fraction as a function of specific angular momentum at different times for both simulations. We find that in the simulation without turbulence (left panel of Figure \ref{fig:angular_pdf}) the gas settles into a triple-peaked distribution by $100$~Myrs (red curve), and the location of these peaks, as well as the overall distribution, does not change over time. In contrast, the simulation with turbulence exhibits an angular momentum distribution with a single peak (right panel of Figure \ref{fig:angular_pdf}). This peak then slowly moves towards lower angular momentum values over time. 

This transport of angular momentum can be explained in terms of the elementary theory of accretion disks \citep[e.g.,][]{Shakura1973A&A....24..337S,Lynden1974MNRAS.168..603L,Pringle1981ARA&A..19..137P}: viscous torques between adjacent annuli of gas in an accretion disk provokes mass to flow inwards. While Gadget2 does contain an artificial viscosity, this treatment is dedicated to capturing shocks in SPH, and has no effect on the transfer of angular momentum, as seen in our turbulence-free simulation (Figure \ref{fig:xy}). The transport of angular momentum is due to ``turbulent'' viscosity induced by our driving method. This explains the slow inward migration of gas in our simulation. 

To estimate the viscosity, $\nu_{\rm turb}$, induced by our turbulence method, we use the accretion rate due to $\alpha$-viscosity \citep{Shakura1973A&A....24..337S,Pringle1981ARA&A..19..137P}:
\begin{equation}
\dot{M} \approx 3\pi \nu_{\rm turb} \Sigma    ,
\end{equation}
where $\dot{M}$ is the mass inflow rate and $\Sigma$ is the surface density. 

\begin{figure}
\centering
\includegraphics[width=.5\textwidth]{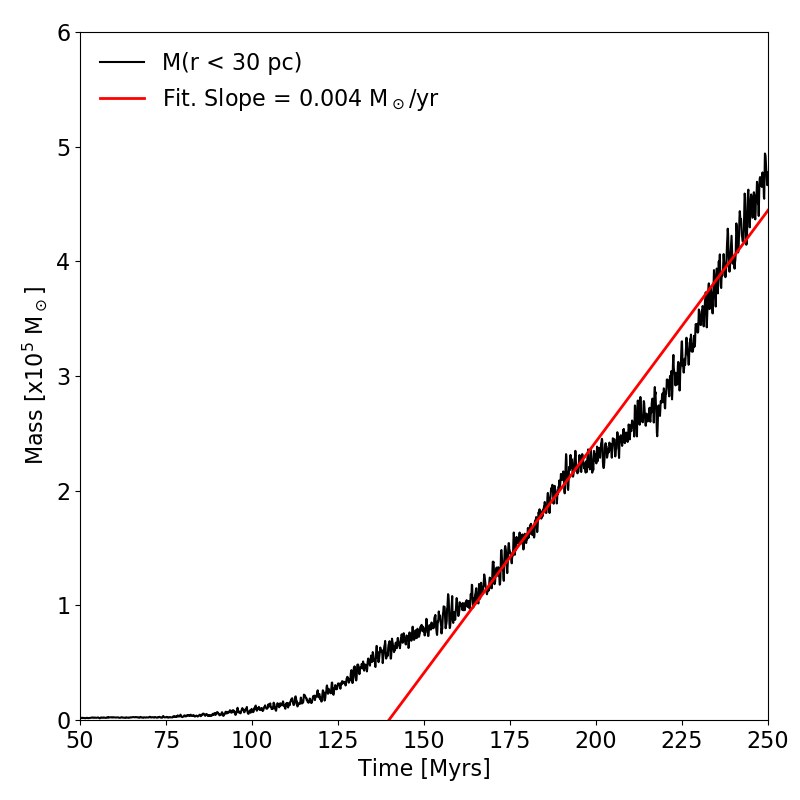}
\caption{Plot of the mass inside a radius of 30 pc over time (simulation with turbulence). To estimate the mass inflow rate, we approximate the rate from t = 150 Myrs to 250 Myrs as a straight line. The slope of the fit is 0.004 \msun/yr.}
\label{fig:mass_in}
\end{figure}

\begin{figure}
\centering
\includegraphics[width=.5\textwidth]{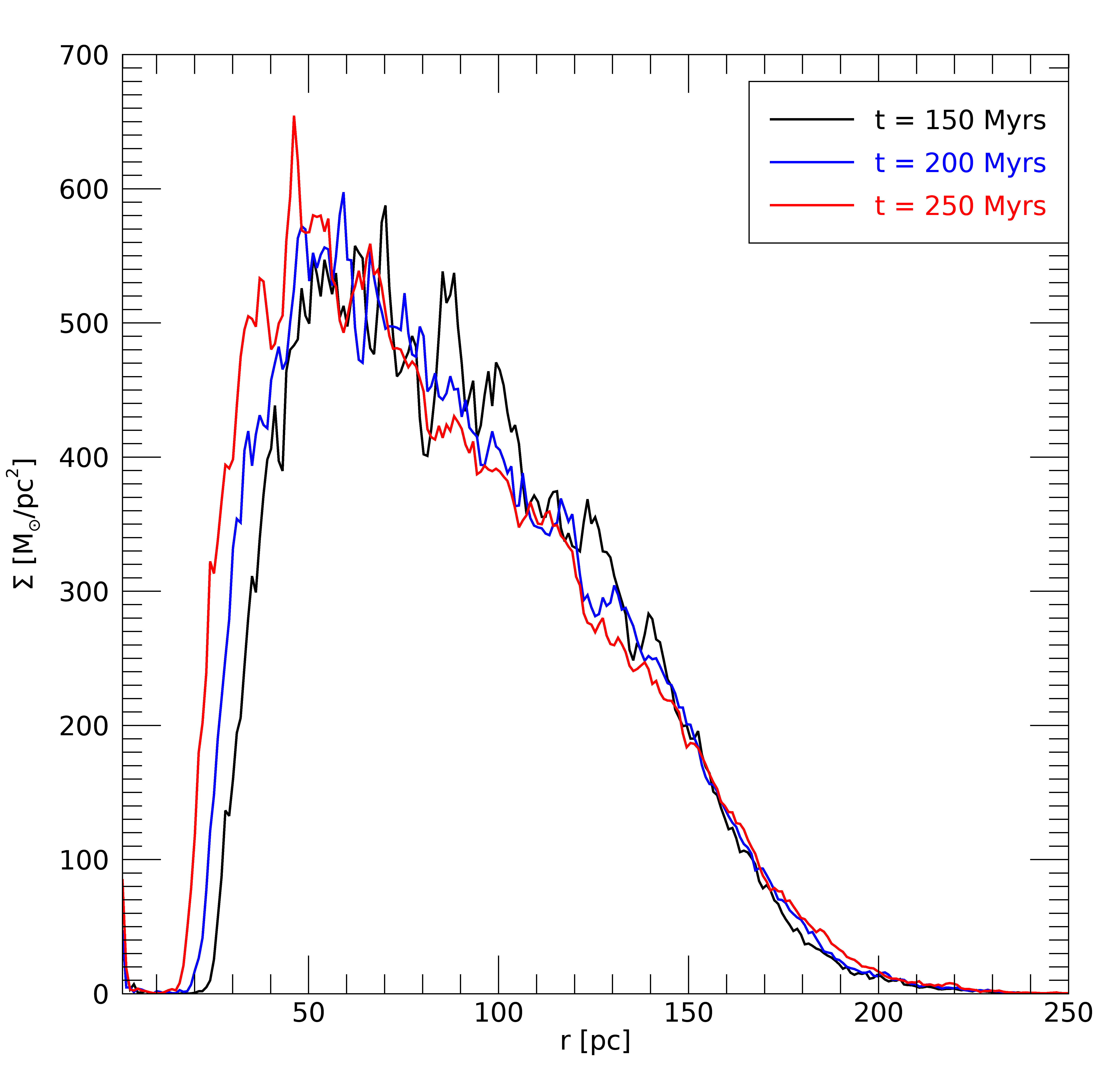}
\caption{Plot of surface density of the simulation with turbulence at t = 150, 200 and 250 Myrs.}
\label{fig:sigma}
\end{figure}

To estimate $\dot{M}$, we calculate the mass inside a radius of 30 pc over time (see Figure \ref{fig:mass_in}). We approximate the rate from 150 Myrs to 250 Myrs as a straight line, and find a slope of $\dot{M}$ = 0.004 \msun/yr. Similarly, we adopt the value of $\Sigma$ at r = 30 pc to be 300 \msun/pc$^2$, which is the average value between 150 and 250 Myrs (see Figure \ref{fig:sigma}). Thus, we evaluate the viscosity to be $\nu_{\rm turb}=4.2\times10^{23}$~cm$^2$/s. 

\begin{figure}
\centering
\includegraphics[width=.5\textwidth]{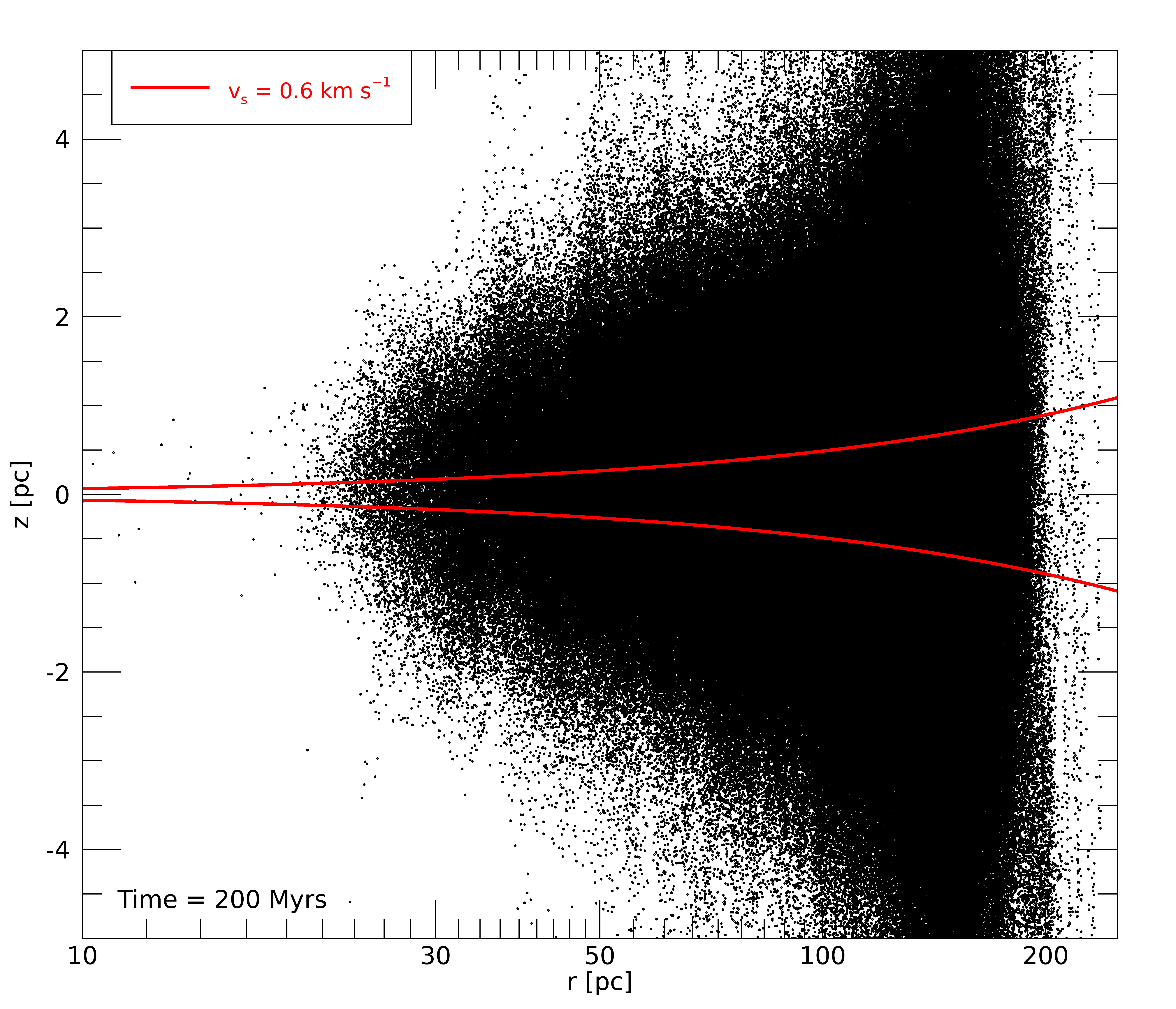}
\caption{Particle position plot z vs. r of the simulation with turbulence at t = 200 Myrs. The red lines indicate the scale height value $H = c_s / \Omega$, where $c_s$ is the sound speed, and $\Omega$ is the angular velocity due to the gravitational potential. Owing to turbulence, the disk is about 10 times thicker than the theoretical value of $H$. }
	\label{fig:scale_height}
\end{figure}

Alternatively, we can use the definition of $\alpha$-viscosity $\nu_{\rm turb} = \alpha c_s H$, where $\alpha \leq 1$ is a parameter that adjusts the strength of the viscosity, $c_s$ is the sound speed, $H = c_s/\Omega$, and $\Omega$ is the angular velocity due to the gravitational potential. Assuming the gas in our simulations is mainly composed of molecular hydrogen ($H_2$), $c_s = 0.64$ km/s. The value of $H$ at r = 30 pc is $\sim0.2$~pc (see Figure \ref{fig:scale_height}). This results in a value for $\nu=3.7\times10^{22}$~cm$^2$/s (assuming $\alpha=1$). However, the actual thickness of the disk at $r = 30$~pc is much larger, $\sim2$~pc (see Figure \ref{fig:scale_height}). Using this value, we obtain $\nu_{turb}=3.7\times10^{23}$~cm$^2$/s, which is comparable to the value we calculated above using the mass inflow rate.

To understand the implications of this calculated turbulent viscosity, we make a comparison similar to that described in \cite{Sormani2018aMNRAS.481....2S}, as follows: to significantly affect the dynamics of a gaseous disk, the viscous timescale, $t_\nu \sim R^2/\nu$ (where R is the radius of the disk), must be shorter than the Hubble time ($t_H$ = 14 Gyrs). This condition gives a lower limit to the value of the viscosity, $\nu_{min} \approx R^2/t_H$. Using $R=200$~pc (the radius of our simulated CMZ disk) gives a minimum viscosity of $\nu_{min} = 8\times10^{23}$~cm$^2$/s. This value  of $\nu_{min}$ is roughly a factor of $2$ higher than the value $\nu_{turb} \approx 4\times10^{23}$~cm$^2$/s calculated from our simulations. This result is consistent with the findings by \cite{Sormani2018aMNRAS.481....2S}, where they found that the viscosity used in their simulations of galactic nuclear rings, $\nu=3\times10^{24}$~cm$^2$/s (which was sufficient to significantly affect the morphology of their simulated rings) was a factor of $\sim10$ smaller than the corresponding minimum viscosity for nuclear rings, $\nu_{min} = 8\times10^{25}$~cm$^2$/s (for a $R=1$~kpc ring). Thus, our results support \cite{Sormani2018aMNRAS.481....2S} conclusion that viscosity may be more effective in influencing the dynamics of gaseous systems than implied by generic estimates such as the one described above.

We note that our viscosity calculations are only approximations, since the values of $\dot{M}$, $\Sigma$, and the thickness of the disk, vary over radius and time. Thus, the viscosity induced by our turbulence method will also vary with radius and time. A more careful analysis of the turbulent viscosity imposed by our driving method is reserved for a future paper.

\section{Summary and Discussion}\label{sec:discussion}
Supersonic turbulence occurs over a wide range of length scales in the interstellar medium, especially within molecular clouds. The importance of turbulence in modulating star formation in the interstellar medium was further highlighted recently by a combination of numerical and analytical studies  \citep[e.g.,][]{Krumholz2005ApJ...630..250K,Burkhart2018ApJ...863..118B}. It has also been suggested that turbulence plays a key role in forming the very first star-clusters and perhaps even globular cluster progenitors \citep[e.g.,][]{Naoz+14,Chiou+19}. Furthermore, turbulence in the centers of galaxies, particularly in our own CMZ, seems to greatly influence its thermal structure and star formation rate \citep[e.g.,][]{Kruijssen2014MNRAS.440.3370K}.

Numerical simulations have shown that turbulence decays quickly, within a few dynamical timescales \citep[e.g.,][]{Stone1998ApJ...508L..99S,MacLow1999}. Thus, in order to properly study the dynamics of the CMZ environment with numerical simulations, it is necessary to include a mechanism for turbulence driving. Typically, simulating turbulence in gas on galactic scales is achieved by modeling star formation and feedback. However, we note that despite the high densities and the large amount of available gas, there is about an order of magnitude less active star  formation in the CMZ than might be expected from the quantity and surface density of molecular gas \citep[e.g.,][]{Longmore2013MNRAS.429..987L,Kruijssen2014MNRAS.440.3370K}. Therefore, other physical processes may supplement this methodology to explain the high-velocity dispersions observed in the CMZ. 

This prompted us to develop a more general mechanism for driving turbulence. Consequently, we adapted the Fourier forcing module described by \cite{MacLow1999} in order to simulate not just the CMZ, but this method can be applied in general to gas that orbits under the influence of a gravitational potential. We implemented this turbulence method to the SPH code Gadget2.

Rather than depending on a single grid of turbulence, as is typical with this method, we create instead many turbulent grids and use them to fill the volume of a larger simulation box. We then use trilinear interpolation to calculate and add a velocity kick to each SPH particle. The amplitude of these velocity increments is adjusted every injection time to maintain a constant energy input. As shown in Section \ref{sec:results}, discontinuities in the added velocity field to the interfaces between grids are indistinguishable in the simulations. 

Using a simplistic model of the CMZ consisting of a collection of Giant Molecular Clouds, our turbulence driving method allows us to study the dynamics of the gas and turbulence over long timescales. One of the main results in our simulations is that turbulence induces inward migration of gas, a result that is consistent with previous numerical simulations \citep[e.g.,][]{Hobbs2011MNRAS.413.2633H}. 

This paper focusses on testing this new method in the context of a single physical scenario. However, our turbulence module is flexible and can be applied to study different physical scales. There are a number of parameters that can be adjusted:
\begin{itemize}
    \item Power spectrum index: For incompressible turbulence, the Kolmogorov power spectrum in three dimensions is $\propto k^{-11/3}$ \citep{Kolmogorov1941DoSSR..30..301K} (for a two-dimensional distribution, $P \propto k^{-8/3}$). For compressible turbulence, the power law index has been shown to be slightly steeper \citep[e.g.,][]{Clark2011}. Milky Way observations have found power law slopes of -2.8 to -3.2 in 2D maps \citep[][and references therein]{Elmegreen2004ARAA..42..211E}. In our tests we used a power spectrum index of -4, but the power spectrum index can be modified to match observations in order to create more applicable simulations. 
    \item The number of distinct realizations of turbulence grids: in order to avoid coherence in scales larger than $L_{cube}$, each of the turbulence grids has a different realization of velocity fields, drawn randomly from a library of 10 realizations. This library can be expanded with more realizations, which could potentially improve the ``randomness'' of the turbulence. However, this library will be limited by the memory constraints of the computing resources.
    \item The size of the turbulence grids ($L_{cube}$): in our code, this parameter also represents the largest scales on which turbulence is injected. Thus, modifying this allows flexibility in studying different turbulence injection scales depending on the physical environments to be simulated. For example, in \cite{Salas2020arXiv201004170S}, we use our driving method to study how turbulence affects the formation of galactic nuclear rings. There, we discuss galactic scale ($L_{global}$ = 4 kpc) simulations using larger grids of turbulence ($L_{cube} = 64$ pc).
    \item The number of turbulence grids per side: modifying this parameter (in tandem with the previous one, $L_{cube}$), allows for adjusting the size of the overall simulation domain, $L_{global}$.
    \item The total energy of injection, \eturb, and time interval between injections, \nt: these parameters need to be tuned depending on the physical environment to be simulated. In Appendix \ref{apen:convergence} we vary these parameters in order to find the optimal values to use. However, different gas configurations will require different parameters than the ones used here. For example, we expect that smaller values of \eturb\ and larger values of \nt\ may be sufficient to balance self-gravity in lower density gas.
\end{itemize}


\acknowledgments
J.M.S would like to thank Paul Clark for his help on adapting the external pressure term into \gadget. Furthermore, J.M.S would like to thank Blakesley Burkhart and Diederik Kruijssen for very helpful discussions about turbulence, and finally, to Sofia G. Gallego and Sungsoo S. Kim for their help in implementing the gravitational potential of the inner Galaxy. This material is based upon work supported by the National Science Foundation Graduate Research Fellowship Program under Grant No. DGE-1144087. SN acknowledges the partial support of NASA grant No. 80NSSC20K0500 and thanks Howard and Astrid Preston for their generous support. This work used computational and storage services associated with the Hoffman2 Shared Cluster provided by UCLA Institute for Digital Research and Education's Research Technology Group. This work also used the Extreme Science and Engineering Discovery Environment (XSEDE) Comet at the San Diego Supercomputer Center at UC San Diego through allocations AST170039 and AST180051. XSEDE is supported by National Science Foundation grant number ACI-1548562.

{\bf Software:}
Figure \ref{fig:ICs}, \ref{fig:xy} and \ref{fig:zx} were done using the SPH visualization software \splash\ \citep{Price2007}. We used Gadget2 \citep{Springel2005} to build our turbulence method. The version of the code that includes our turbulence routine can be found at \href{https://github.com/jesusms007/turbulence}{https://github.com/jesusms007/turbulence}.

\appendix

\section{Constant energy input rate}
\label{apen:derivation}
In this appendix we derive Equation \ref{eq:A}. 

Our algorithm adds a velocity ``kick'' to each particle every $N_t$ timesteps while  maintaining a constant energy input rate $\dot{E_{in}} = \Delta E_{in} / (N_t \Delta t$), where $\Delta t$ is the simulation timestep (fixed to be equal to 1000 yrs), and:
\begin{equation} \label{eq:A1}
\Delta E_{in} = E_2 - E_1 = \frac{1}{2} m_p \sum_{i=1}^{N_{sph}} \vec{v_{2,i}} \cdot \vec{v_{2,i}} - \frac{1}{2} m_p \sum_{i=1}^{N_{sph}} \vec{v_{1,i}} \cdot \vec{v_{1,i}} \ ,
\end{equation}
where $\vec{v_{1,i}}$ is the velocity vector of a particle at time $t_1$ (before the kick), and
\begin{equation}
\label{kick}
\vec{v_{2,i}} = \vec{v_{1,i}} + A\Delta \vec{v_i} = \vec{v_{1,i}} + A \sqrt{G \rho_i} \text{     } \vec{I_i}(x,y,z)
\end{equation}
is the velocity of the particle at time $t_2$ (after the kick). Therefore, $t_2 - t_1$ = $N_t\Delta t_s$. The function $\vec{I}(x,y,z)$ is the interpolation function that represents the turbulent velocity increment based on the particle's position (hereafter called $\vec{I}$), and $A$ is the target variable.

Equation \ref{eq:A1} then becomes:
\begin{equation} \label{eq:A1_cont}
\Delta E_{in}  = \frac{1}{2} m_p \left[ \sum_{i=1}^{N_{sph}} \left(\vec{v_{1,i}} + A \sqrt{G \rho_i} \text{     }  \vec{I_i} \right) \cdot \left(\vec{v_{1,i}} + A \sqrt{G \rho_i} \text{     }  \vec{I_i} \right) -  \sum_{i=1}^{N} \vec{v_{1,i}} \cdot \vec{v_{1,i}}  \right]
\end{equation}

By simplifying Equation \ref{eq:A1_cont}, the result is:
\begin{equation}
\label{eq:A_deriv}
\Delta E_{in} = \frac{1}{2} m_p \sum_{i=1}^{N_{sph}} \left[      A^2  G \rho_i \vec{I_i} \cdot \vec{I_i}  + 2 A \sqrt{G \rho_i} \text{     }  \vec{I_i} \cdot \vec{v_{1,i}}       \right]
\end{equation}

This equation always gives both positive and negative values for $A$ and our code always chooses the positive value. However, to ensure that the injection of turbulence does not violate the conservation of total angular momentum of the gas, we simply multiply $A$ by a factor of $-1$ or $+1$, alternating between these two factors every time the driving is performed. This ensures that, over time, the net angular momentum added to the gas particles is approximately zero while keeping the same increase in energy $\Delta E_{in}$. 

Furthermore, we change each of the $64^3$ cubes of turbulence every time the driving is performed. This is to mimic the random nature of turbulence. Regardless of the source of turbulence, we expect that a parcel of gas will experience a coherent turbulent driving force during a given time interval. We can justify changing each grid every $N_t$ timesteps if we consider that the crossing time of a parcel of gas traveling at an orbital speed of v$\approx150$ km s$^{-1}$ through the average scale of a turbulence cube, i.e., $L_{ave}\approx1$~pc, is $t_{cross}\approx6000$~yrs. Hence, we conclude that there is little need for the velocity field to be coherent on timescales longer than 6 timesteps, thus justifying our replacement of each turbulence cube as long as $N_t$ is more than a few.
\section{Dependence on turbulence parameters}\label{apen:convergence}
Here we describe the tests we carried out to study the performance of our turbulence algorithm in order to choose \nt\ and \eturb\ values for our CMZ model. While these parameters are better represented in form of a rate, the results of this section indicate that the time interval at which the turbulence is injected affects the subsequent evolution of the gas. 

We tested driving the turbulence every 1 to 5 timesteps for a total of 25 tests: for each \nt, we used \eturb\ = $10^{46}$ to $10^{50}$ ergs, in a factor of 10 increments. We ran each simulation for 100 Myrs to allow the system to reach a quasi-steady state. 

\begin{figure}
	\centering
	\includegraphics[width=\textwidth]{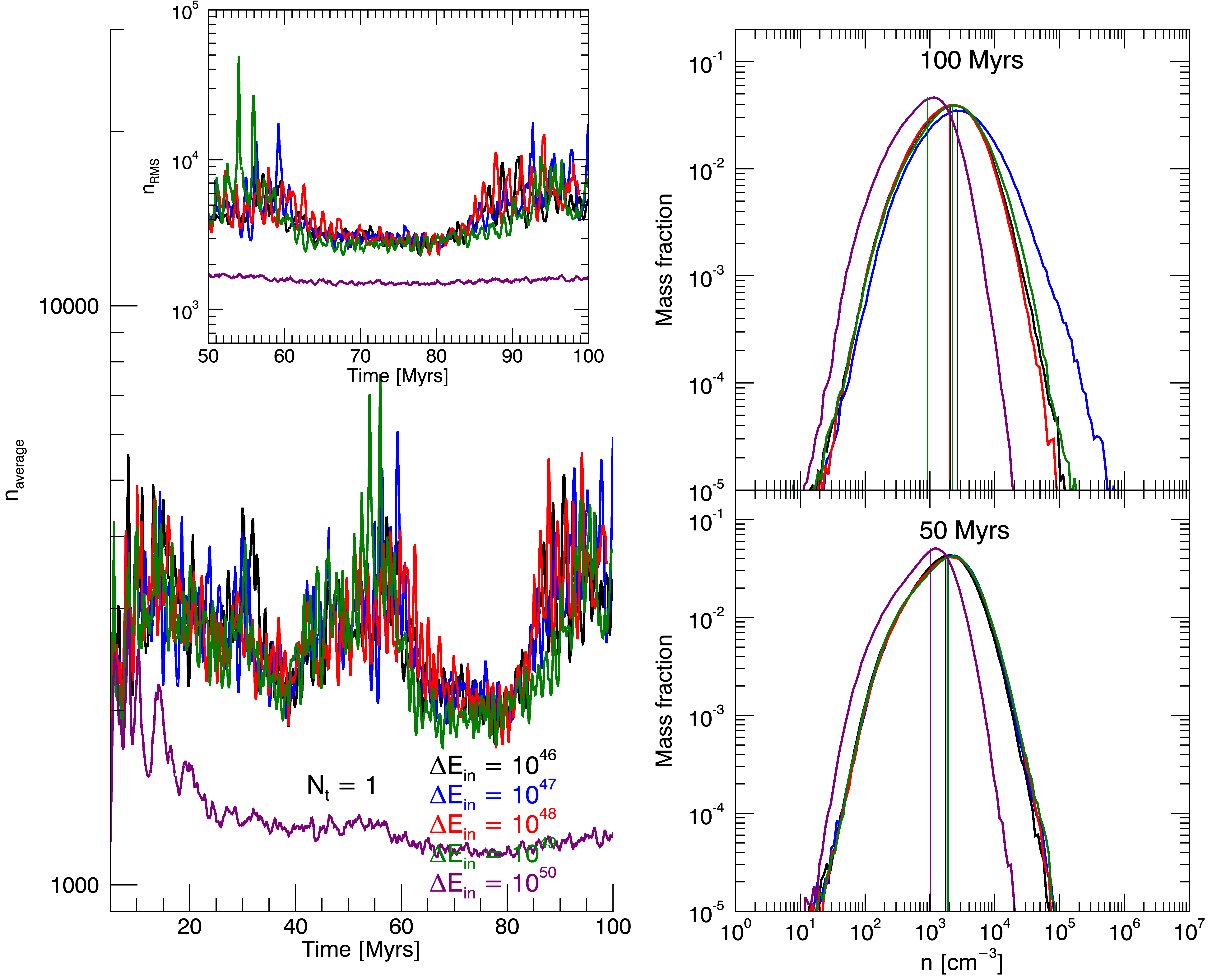}
	\caption{Median density, RMS density and density PDF for tests with \nt\ = 1. The mass-weighted median density is plotted with a vertical line.  The density PDF is approximately lognormal. }
	\label{fig:convergence1}
\end{figure}

\begin{figure}
\centering
\includegraphics[width=\textwidth]{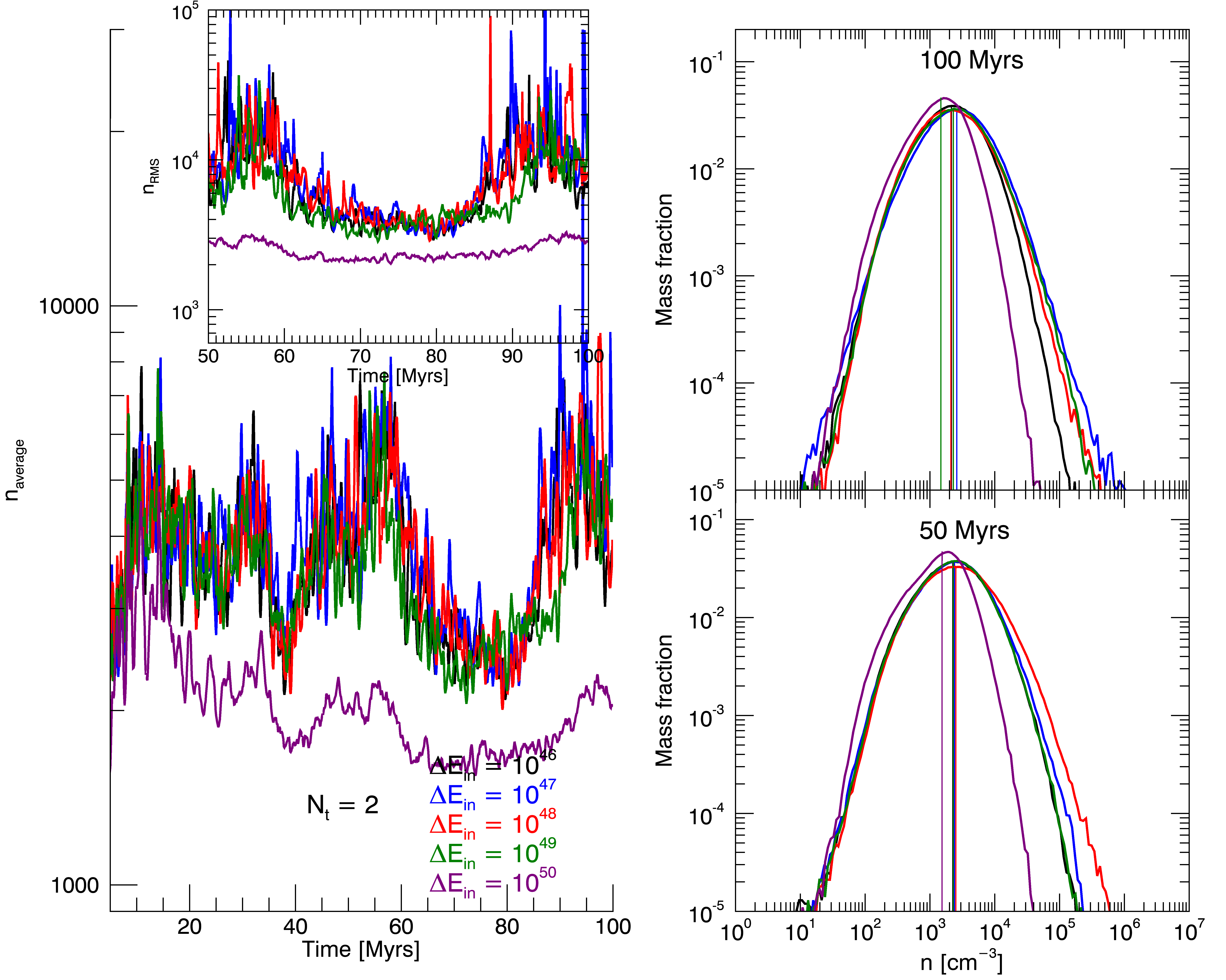}
\caption{Median density, RMS density and density distribution function for tests with \nt\ = 2. The distribution is lognormal. The mass-weighted median density is plotted with a vertical line.  }
\label{fig:convergence2}
\end{figure}

\begin{figure}
\centering
\includegraphics[width=\textwidth]{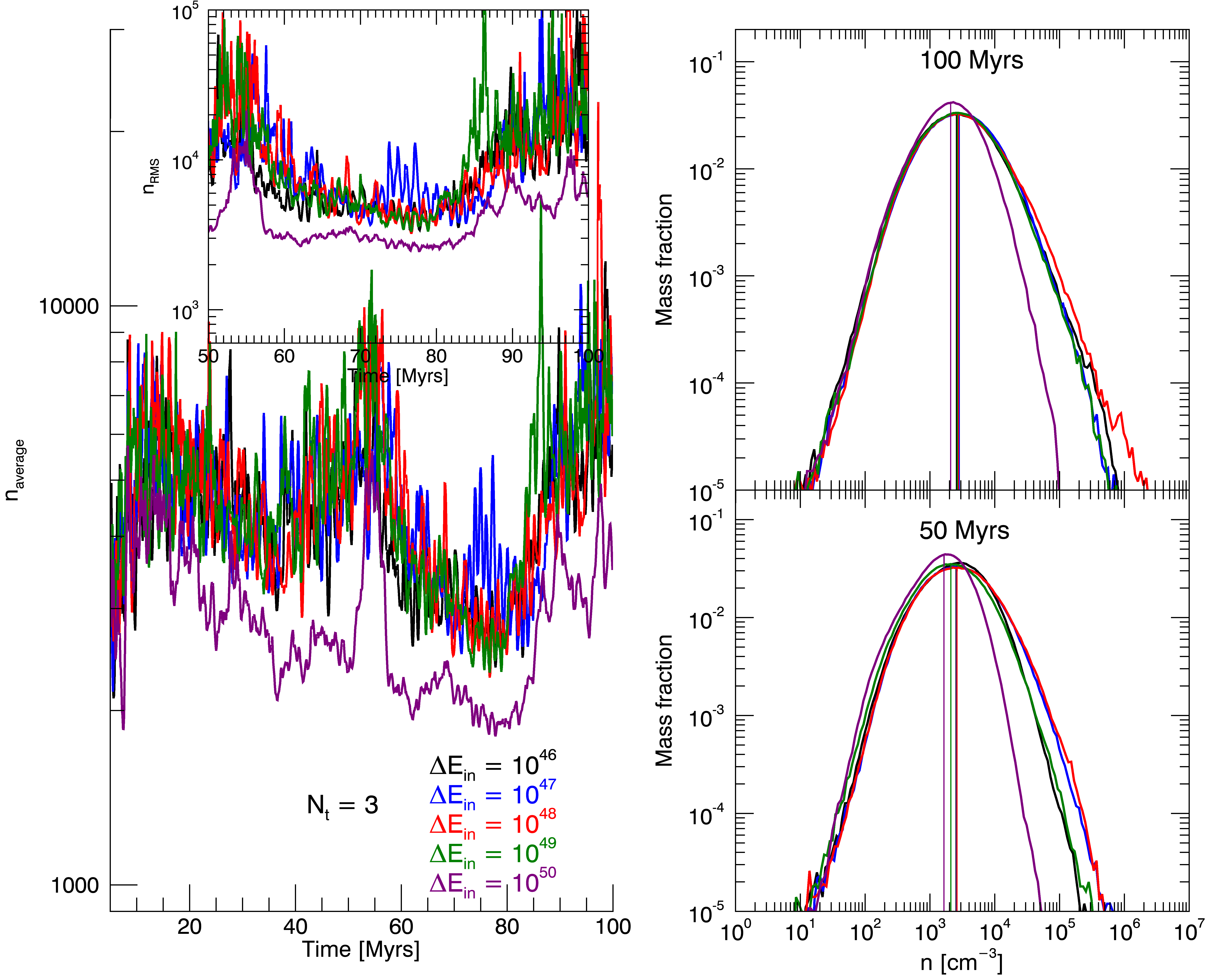}
\caption{Median density, RMS density and density distribution function for tests with \nt\ = 3. The distribution is lognormal. The mass-weighted median density is plotted with a vertical line.   }
\label{fig:convergence3}
\end{figure}

\begin{figure}
\centering
\includegraphics[width=\textwidth]{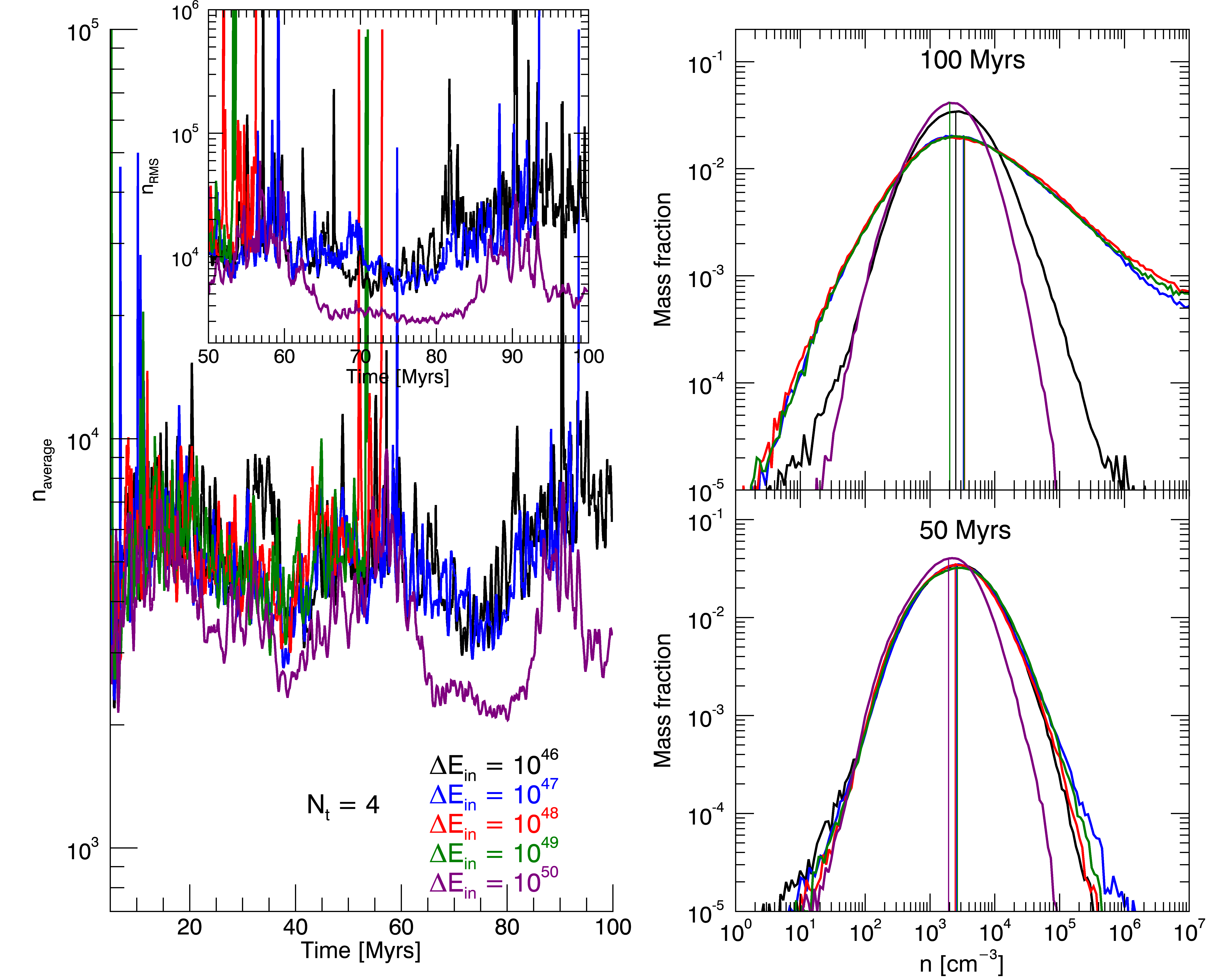}
\caption{Median density, RMS density and density distribution function for tests with \nt\ = 4. The distribution is lognormal. The mass-weighted median density is plotted with a vertical line. }
\label{fig:convergence4}
\end{figure}

\begin{figure}
\centering
\includegraphics[width=0.95\textwidth]{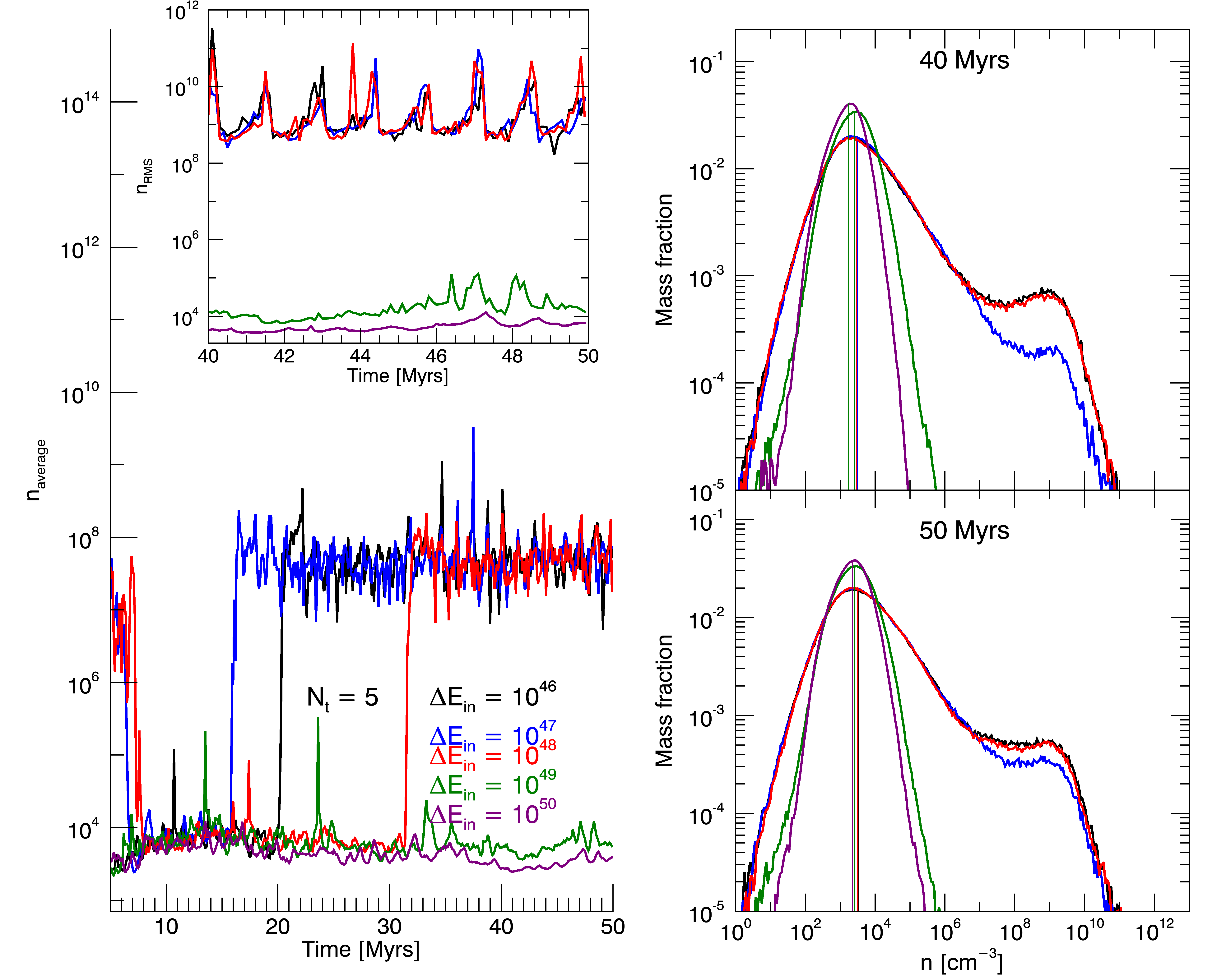}
\caption{Median density, RMS density and density distribution function for tests with \nt\ = 5. The mass-weighted median density is plotted with a vertical line. These tests were run only up to 50 Myrs. The density PDF of the runs with \eturb\ = $10^{46}-10^{48}$ ergs deviate dramatically from lognormal.  }
\label{fig:convergence5}
\end{figure}

As proxy for the evolution of the systems, we plot the average number density ($n_{ave}$) as a function of time, as shown in Figures \ref{fig:convergence1} to \ref{fig:convergence5}. For the simulations with \nt = 2 and 3, the time evolution of $n_{ave}$ is very similar (except for the case with \eturb = $10^{50}$ ergs, whose $n_{ave}$ evolution diverges from all other runs). The average density oscillates due to the interplay between turbulence and self-gravity, potentially reaching a steady-state by $\sim50$ Myrs.

We also plot in Figures \ref{fig:convergence1} to \ref{fig:convergence5} the RMS number density ($n_{RMS}$) from $t$ = $50-100$ Myrs, in order to better discern differences between runs. The runs with parameter \nt = 4 and 5 show higher average densities, as expected, because in these cases, turbulence is injected less frequently, allowing gravity more time to compress the gas to higher densities. However, given their high density RMS peaks, these runs are less consistent and more chaotic than those with lower values of \nt. Also, the runs with \nt\ = 5 were only run for 50 Myrs due to the fact that the tests with \eturb\ = $10^{46}$ to $10^{48}$ ergs exhibited high density clumps which slowed down the computation time. These clumps originate because in those cases, the turbulence is not injected often enough to support the gas against local gravitational collapse. Particles pile on top of each other, and due to the nature of the kernel used by \gadget, once the distance between particles approaches the smoothing length, the pressure gradient is no longer correct and the particles stick together, creating very high density clumps.

Additionally, in Figures \ref{fig:convergence1} to \ref{fig:convergence5}, we plot the distribution of mass fraction as a function of density (density PDF) at $t = 50$ and $100$ Myrs for each run. Inside a molecular cloud, the density PDF is shaped by the complex interaction between turbulence, self-gravity, magnetic fields and stellar feedback. As a result, it is an effective tool to determine the dynamical state of the gas (e.g. \citealt{Federrath2010AA...512A..81F}). In most of our test runs, the calculated density PDF resembles a lognormal distribution, a result that several groups have found for isothermal gas (e.g., \citealt{Vazquez-Semadeni1994ApJ...423..681V,Klessen2000ApJ...535..869K} and others). This gives us confidence that our new turbulence forcing module is consistent with previous studies of turbulence. Thus, we can fit the density PDFs to Gaussian functions of $x$ = $\ln(n/n_0$) with mean $\mu$ and dispersion $\sigma$:
\begin{equation}\label{eq:gaussian}
f_M = C \text{ exp}\left[ \frac{(x - |\mu|)^2}{\sigma^2}\right] .
\end{equation}
The mass-weighted median number density (half of the mass is at densities above and below this value) is proportional to $e^\mu$. We indicate this parameter with vertical bars in both figures. For each value of \nt\ = 1, 2 and 3, the value mass-weighted median number density is consistent for all energies, except \eturb\ = $10^{50}$.

Finally, it is a standard result in the literature that high-density regions are created by turbulence due to supersonic turbulent convergent flows (e.g., \citealt{Elmegreen2004ARAA..42..211E,McKee2007ARAA..45..565M}). The increasing values of $n_{ave}$ with \eturb\ seen in Figures \ref{fig:convergence1} to  \ref{fig:convergence4} agrees with this result.

The tests with parameter \nt\ = 2 and 3 show better consistency in the range of densities in the simulations ($n\sim$ 10$^{3-4}$ \invcmcube) than the other test runs. Based on their density evolution, we notice that the exact choice of of energies in this range of injection intervals has little effect on the final results. Thus we opt for the parameters \nt\ = 2 and \eturb\ = $10^{47}$ ergs as the standard choice for our CMZ model simulated in Section \ref{sec:results}. 

As a final note, we remind the reader that we are only testing this parameter-space of injection energies and intervals in a specific physical scenario, namely, that of a CMZ-like environment. We expect that similar choice of parameters will be useful in different environments, but further testing is needed. However, our goal here is to show that our proof-of-concept method gives consistent results and agrees with previous results in the literature. A more comprehensive analysis of the turbulence method that we introduce in this paper is left for future work. 

\clearpage 

\bibliographystyle{aasjournal}
\bibliography{references}

\end{document}